# Precessing equilibrium and overcritical solitons at a spin-flop phase transition

## V.V.Nietz

*Frank and Shapiro Laboratory of Neutron Physics, Joint Institute for Nuclear Research, Dubna, Moscow Region 141980, Russia*

## Abstract

Taking into account the energy dissipation in the spin-flop phase transition induced by magnetic field in an antiferromagnet with uni-axial anisotropy, the following peculiarities of precessing ball solitons (PBS) are considered: a) the states of equilibrium precessing solitons arising in the initial phase during the phase transition; b) the states of "overcritical PBS" existing outside the region of metastability. The "overcritical" PBS may originate during disintegration of the initial phase.

Tel: +7-496 21-65-552
Fax: +7-496-21-65-882
*E-mail address*: nietz@nf.jinr.ru

## Introduction

The main properties of precessing ball solitons (PBS) in an antiferromagnet at the spin-flop transition, induced by the external magnetic field, have been analyzed in the [1–3] papers. In the given paper, in addition to these papers, we consider the so-called “equilibrium PBS” that are possible even taking into account the dissipation for precessing magnetic moments. Besides, the PBS in the overcritical range of a field, i.e. outside the metastability region, are considered.

## 1. The equations for PBS

To analyze magnetic solitons in an antiferromagnet, the expression for the macroscopic energy

$$W = 2M_0\int\left\{-\frac{A}{2}|\mathbf{l}|^2+\frac{B}{2}|\mathbf{m}|^2+\frac{C}{4}(\mathbf{ml})^2+\frac{K_1}{2}\left(|m_\perp|^2+|l_\perp|^2\right)-\frac{K_2}{4}\left(|m_\perp|^2+|l_\perp|^2\right)^2-m_zH_z+\right.$$
$$\left.+\frac{\alpha_{xy}}{2}\left[\left(\frac{\partial\mathbf{m}}{\partial X}\right)^2+\left(\frac{\partial\mathbf{m}}{\partial Y}\right)^2+\left(\frac{\partial\mathbf{l}}{\partial X}\right)^2+\left(\frac{\partial\mathbf{l}}{\partial Y}\right)^2\right]+\frac{\alpha_z}{2}\left[\left(\frac{\partial\mathbf{m}}{\partial Z}\right)^2+\left(\frac{\partial\mathbf{l}}{\partial Z}\right)^2\right]\right\}dXdYdZ \quad (1)$$

and the equations of motion

$$\frac{M_0\hbar}{\mu_B}\frac{\partial\mathbf{l}}{\partial t}=\mathbf{m}\times\frac{\delta W}{\delta\mathbf{l}}+\mathbf{l}\times\frac{\delta W}{\delta\mathbf{m}}+\Gamma\frac{M_0\hbar}{\mu_B}\left(\mathbf{m}\times\frac{\partial\mathbf{l}}{\partial t}+\mathbf{l}\times\frac{\partial\mathbf{m}}{\partial t}\right), \quad (2)$$

$$\frac{M_0\hbar}{\mu_B}\frac{\partial\mathbf{m}}{\partial t}=\mathbf{m}\times\frac{\delta W}{\delta\mathbf{m}}+\mathbf{l}\times\frac{\delta W}{\delta\mathbf{l}}+\Gamma\frac{M_0\hbar}{\mu_B}\left(\mathbf{m}\times\frac{\partial\mathbf{m}}{\partial t}+\mathbf{l}\times\frac{\partial\mathbf{l}}{\partial t}\right) \quad (3)$$

are used [3].

Here **m** and **l** are non-dimensional ferromagnetic and antiferromagnetic vectors; $l_\perp = l_x + il_y, m_\perp = m_x + im_y$; the absolute value of the vector **l** at $H=0$ equals $1$, $M_0$ is the magnetization of each sublattice.

The solutions of equations (2) and (3) can be presented in the following form:

$$l_\perp(\mathbf{r},\tau)=q(\mathbf{r},\tau)e^{i(\omega(\tau)\tau-k(\tau)x)},\quad m_\perp(\mathbf{r},\tau)=p(\mathbf{r},\tau)e^{i(\omega(\tau)\tau-k(\tau)x)}. \quad (4)$$

In the case of immovable PBS, i.e. at $k\equiv 0$, the (2), (3) equations have the solutions with a spherical symmetry. For such case, the equation has the following form:

$$\frac{d^2q}{dr^2}+\frac{2}{r}\frac{dq}{dr}+\frac{q}{1-q^2}\left(\frac{dq}{dr}\right)^2=q\left(1-q^2\right)\left(1-(\omega+h)^2-\frac{k_2}{k_1}q^2\right), \quad (5)$$

or the same for $l_z$ parameter:

$$\frac{d^2 l_z}{dr^2}+\frac{2}{r}\frac{dl_z}{dr}+\frac{l_z}{1-l_z^2}\left(\frac{dl_z}{dr}\right)^2=\left((\omega+h)^2-1+\frac{k_2}{k_1}\right)l_z-\left((\omega+h)^2-1+\frac{2k_2}{k_1}\right)l_z^3+\frac{k_2}{k_1}l_z^5 . \qquad (6)$$

This equation should be supplemented by the following expression describing the evolution of PBS:

$$\frac{\partial q}{\partial \tau}\cong -Qk_1\omega(\omega+h)q\left(2q^2-1\right). \qquad (7)$$

In Eqs. (5)–(7), the frequency $\omega$ depends on time.

At $k\equiv 0$, the energy of PBS can be expressed as follows:

$$E_s=8\pi M_0\alpha_{xy}\sqrt{\frac{\alpha_z}{k_1 B}}\int_0^\infty\left\{\left[\frac{1+(\omega+h)^2}{2}-(\omega+h)h\right]q^2-\frac{k_2}{4k_1}q^4+\frac{1}{2\left(1-q^2\right)}\left(\frac{dq}{dr}\right)^2\right\}r^2dr . \qquad (8)$$

In the (5)–(8) expressions, the following designations are used: $x=K_1^{0.5}\alpha_{xy}^{-0.5}X$, $y=K_1^{0.5}\alpha_{xy}^{-0.5}Y$, $z=K_1^{0.5}\alpha_z^{-0.5}Z$; $k_1=K_1/B$, $k_2=K_2/B$, $h=H_z\left(BK_1\right)^{-0.5}$, $R=2M_0B\alpha_{xy}\alpha_z^{0.5}K^{-1.5}$, $\tau=2\mu_B\left(K_1B\right)^{0.5}\hbar^{-1}t$, $Q=\Gamma 2\mu_B k_1^{-0.5}\hbar^{-1}$.

In all subsequent examples, the following parameters, that are typical for antiferromagnetic crystals, have been used: $M_0=0.33\times10^{-9}\,eVOe^{-1}\mathring{A}^{-3}$, $B=4.9\times10^6\,Oe$, $\alpha_{xy}=\alpha_z=3\times10^6\,Oe\,\mathring{A}^2$, $K_1=700\,Oe$, $K_2=140\,Oe$, $Q=0.02$ [4].

## 2. Equilibrium ball solitons

At $h<1$, the curves of $E_s(\omega)$ have a minimum, which is obvious from Fig. 1. Let's consider in detail the regions of the minima only in two cases: at $h=0.99$ in Fig. 2 and $h=0.999$ in Fig. 3.

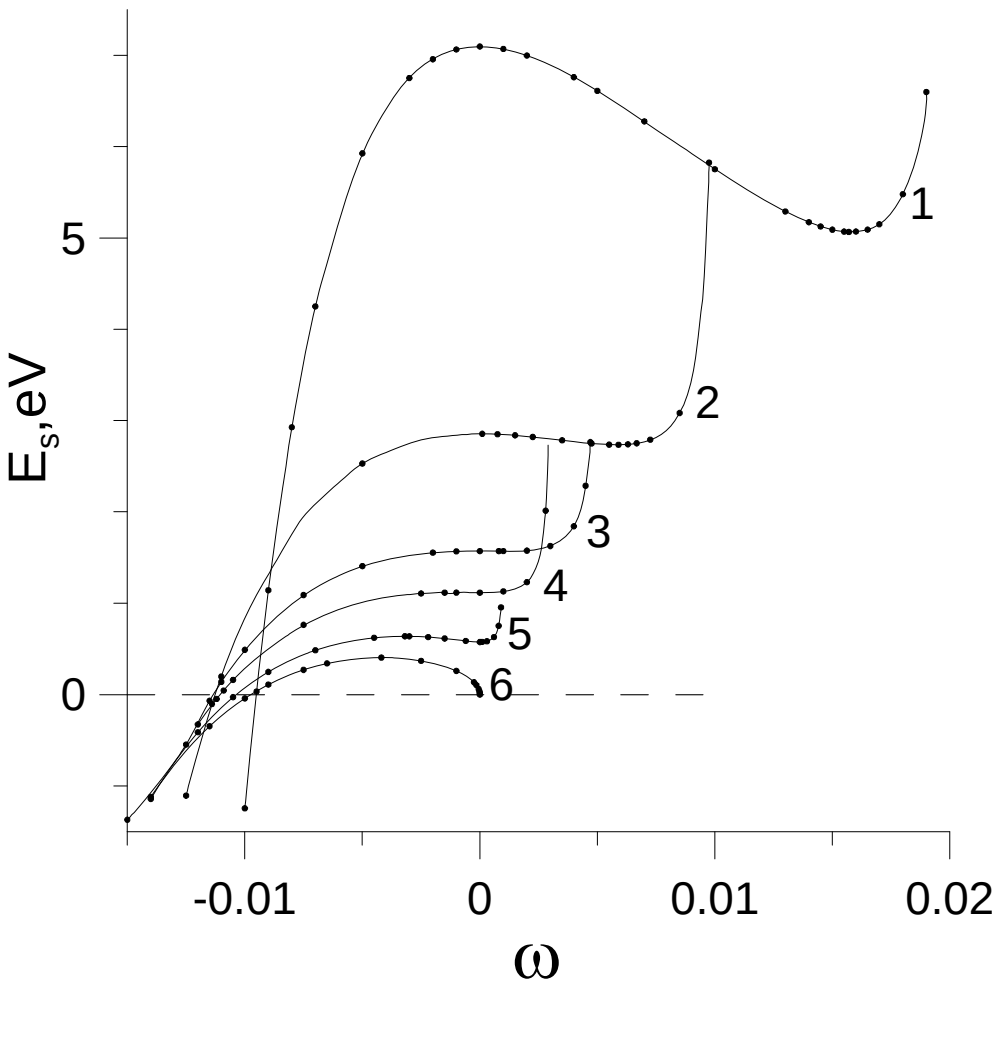

Fig.1 The frequency dependencies of PBS energy for the following values of a field: (1) — 0.98; (2) — 0.99; (3) — 0.995; (4) — 0.997; (5) — 0.999; (6) — 1.

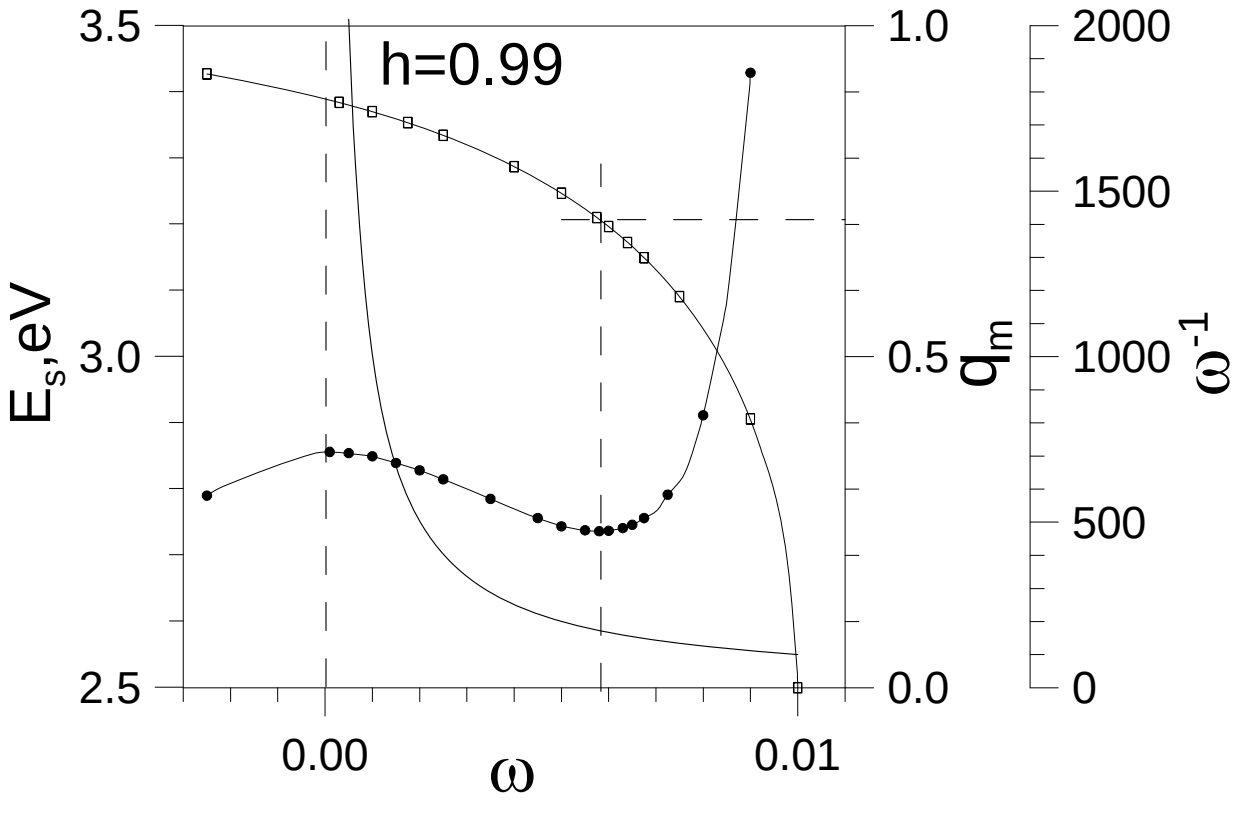

Fig.2 Frequency dependencies of energy, amplitude, and $\omega^{-1}$ value for $h=0.99$. Here and in subsequent figures, the values of energy are denoted by circles; the amplitude, by empty squares; the frequency and the inverse of the frequency, by a continuous line or full squares; the radius, by crosses.

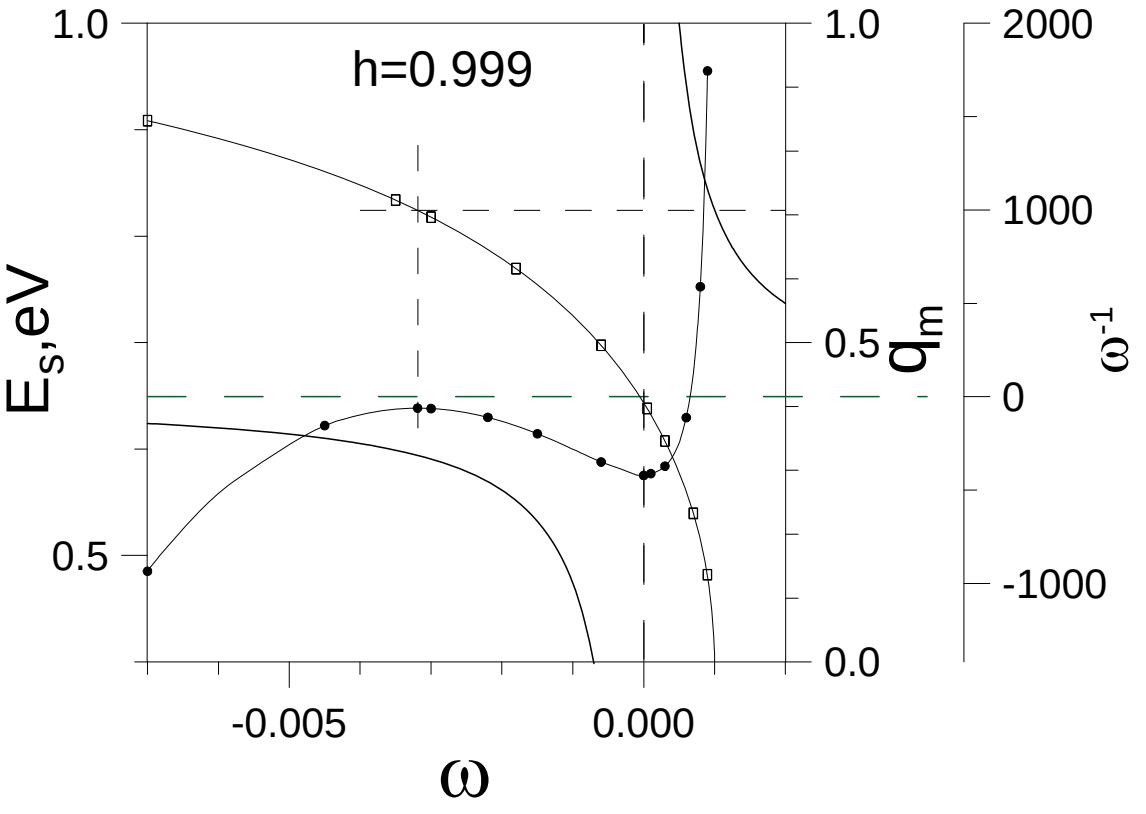

Fig.3 Frequency dependencies of energy, amplitude, and $\omega^{-1}$ value for $h=0.999$.

In the first case, the energy minimum is at $\omega \cong 0.0058, q_m = \sqrt{0.5}$, and the energy maximum is at $\omega = 0$. In the second case, contrary, the minimum is at $\omega = 0$ and the maximum at $q_m = \sqrt{0.5}$.

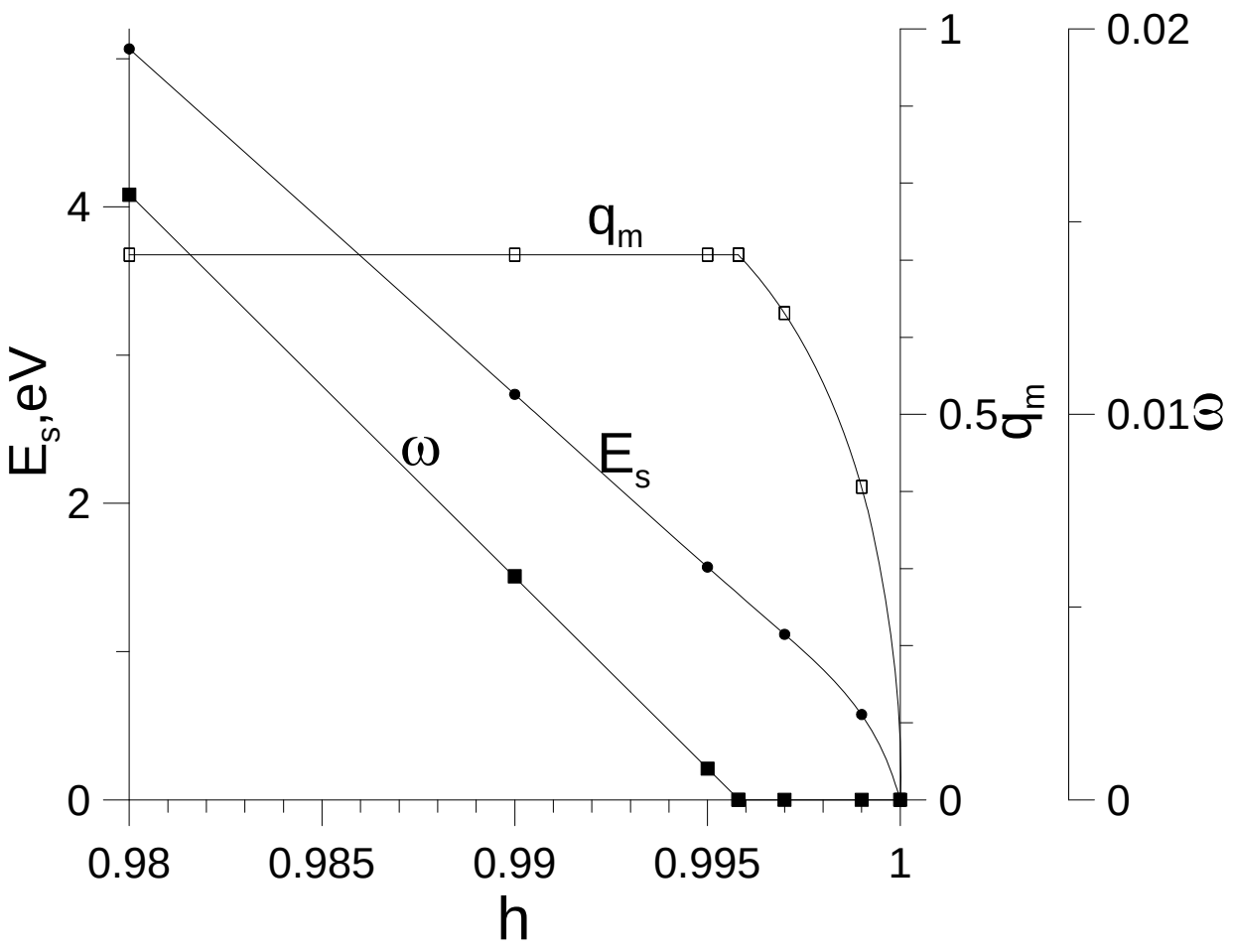

Fig.4 The field dependencies of the energy, amplitude, and frequency for the equilibrium PBS.

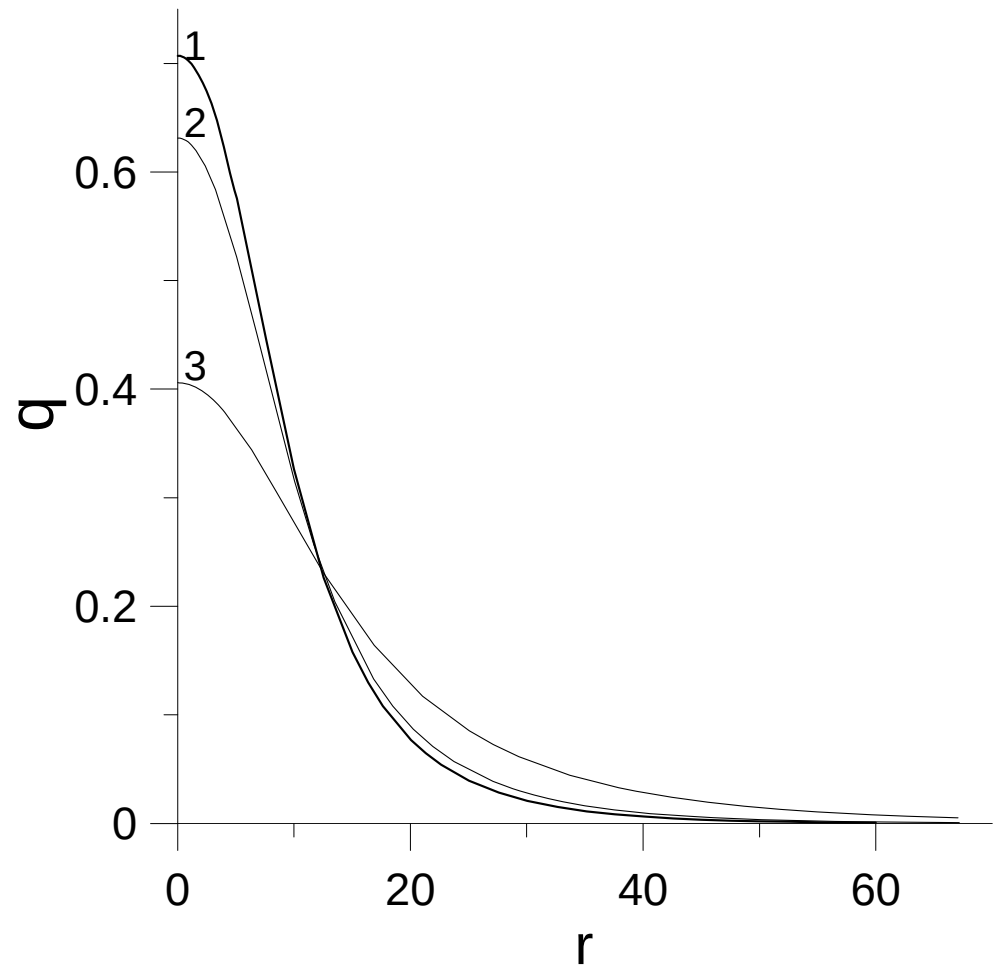

Fig.5 Configurations of the equilibrium PBS: (1) $q_m = \sqrt{0.5}$, (2) $q_m = 0.631$, $h = 0.997$, (3) $q_m = 0.406$, $h = 0.999$.

In Fig. 4, the field dependencies of the energy, amplitude, and frequency for the equilibrium PBS are shown. In Fig. 5, the configurations of equilibrium solitons are presented.

Thus, at $h < 1$ we have two types of equilibrium solitons: precessing equilibrium ball solitons of the first type (EBS-1) and non-precessing equilibrium solitons of the second type (EBS-2). EBS-1 have a fixed amplitude equal to $q_m = \sqrt{0.5}$ and the configuration that does not depend on the field values. The amplitude and configuration of EBS-2, at $\omega = 0$, depend on the magnetic field. The states of EBS-1 exist at $h < h^*$, in our examples $h^* \cong 0.996$. The states of EBS-2 correspond to the condition $1 > h > h^*$.

Let's assume that at the moment $\tau = 0$ the PBS of the $q(r,0)$ form, corresponding to Eq. (5), have arisen. The process of subsequent change of the PBS configuration, in accordance with the expression (7), can be obtained using the procedure used in [3]. Asymptotical approach of precession frequency to equilibrium values as the time function to $\omega \cong 0.0058$ in the first case and to $\omega = 0$ in the second case, on the side of lesser and greater values, are shown in Figs. 6 and 7.

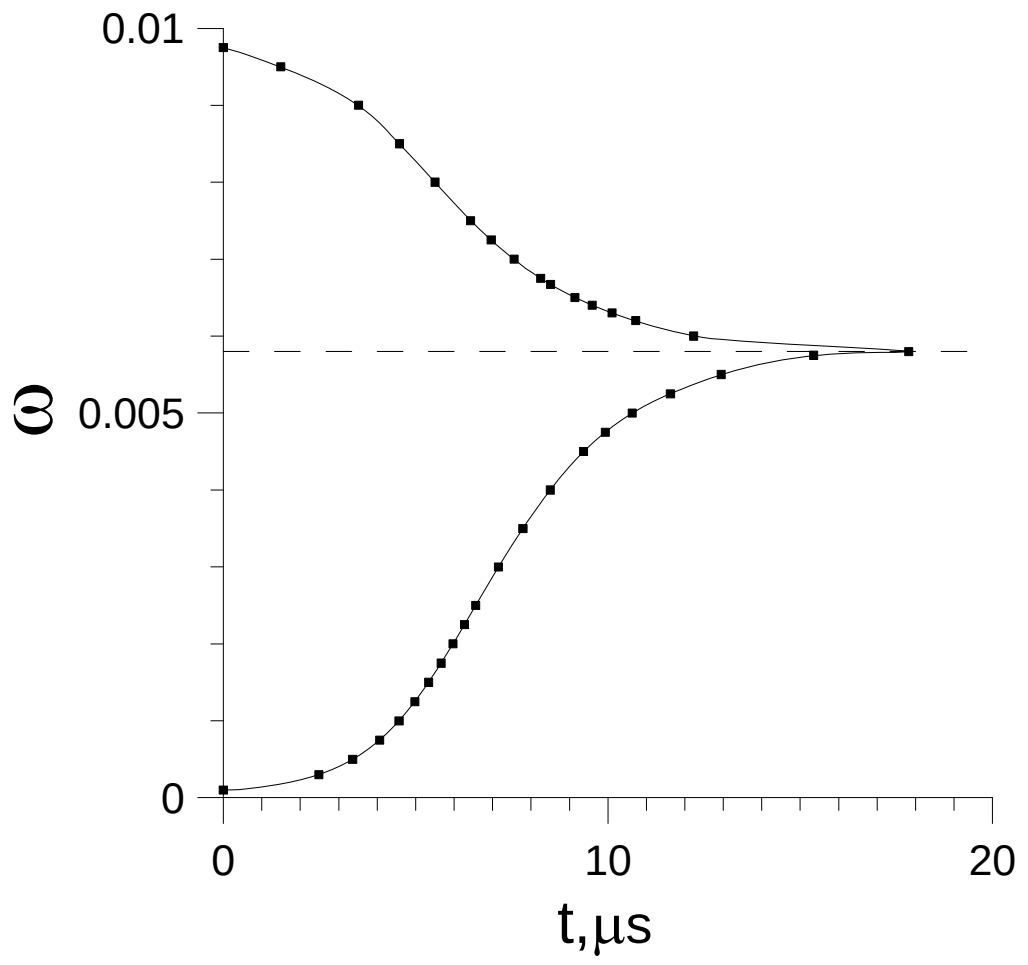

Fig.6 The change of the frequency at approaching of PBS to the equilibrium from the side of lesser and greater values of the frequency if $h = 0.99$. Equilibrium frequency is $\omega_{equal} \cong 0.0058$.

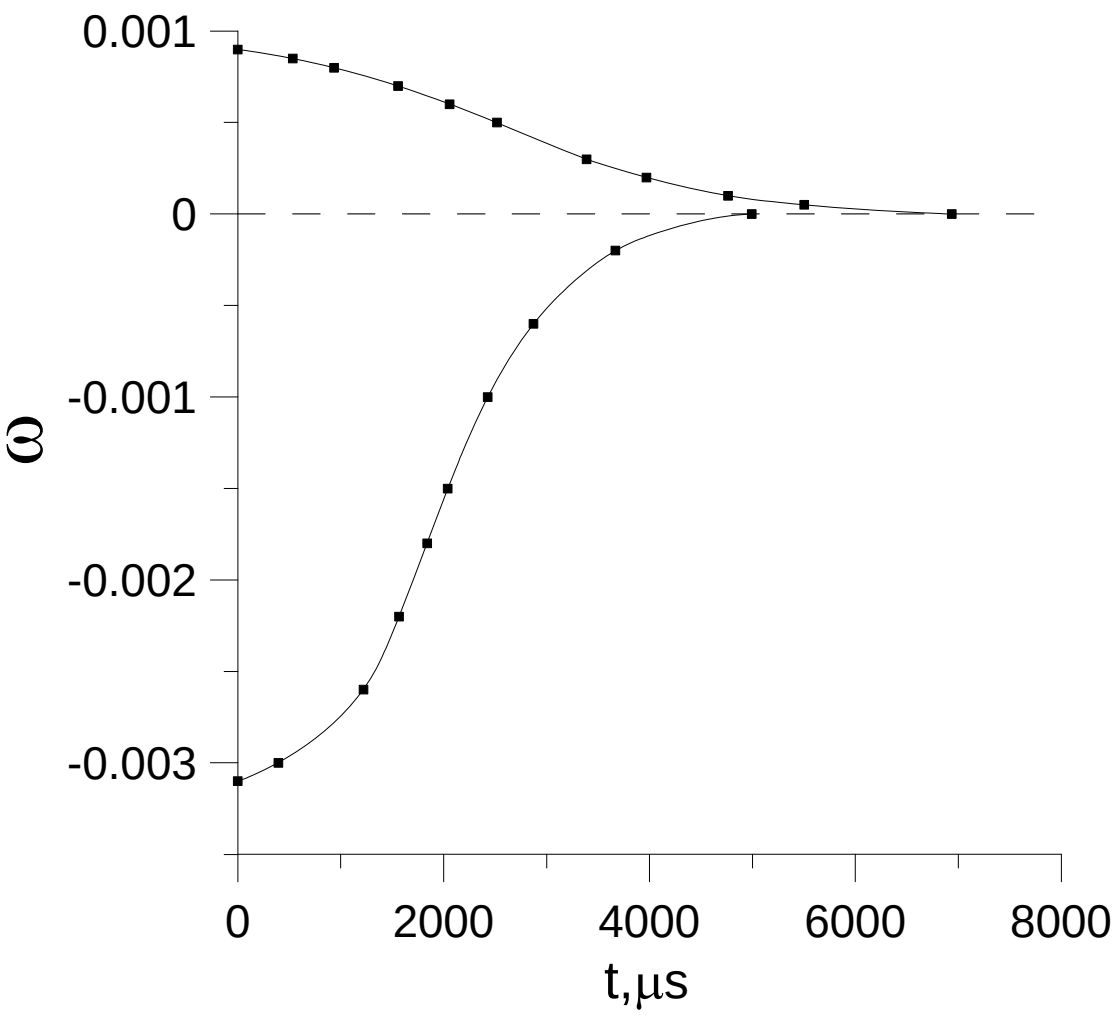

Fig.7 The change of the frequency at approaching of PBS to the equilibrium from the side of lesser and greater values of the frequency if $h = 0.999$.

## 3. Overcritical PBS

The PBS states also exist when the initial-phase state is absolutely unstable [1-3]. Therefore, creation of PBS is possible at disintegration of the initial phase, i.e. at $h > 1$ (and at $h < \sqrt{1 - k_2/k_1}$ for reverse spin-flop transition). It is visible in Figs. 8 and 9. Besides, there is a range of values of a magnetic field where the energy of such PBS can be positive. As seen from Fig. 9, for example, at $h = 1.001$ the creation of PBS with near-zero energy is possible for two various frequencies of precession. It can be seen from Eq. (5) that for all PBS at $h > 1$ the precession frequencies are negative, but at $h < \sqrt{1 - k_2/k_1}$ for all PBS at reverse transition the precession frequencies are positive.

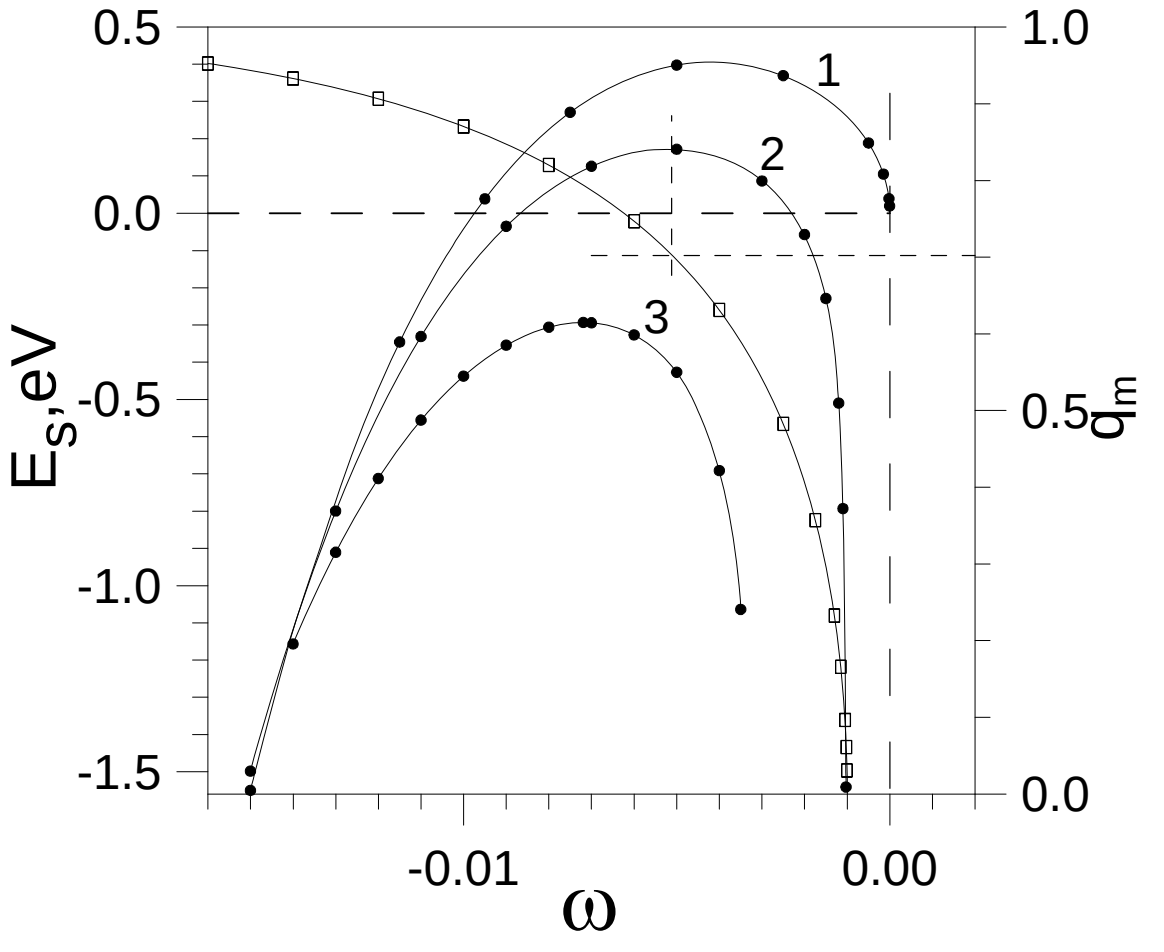

Fig.8 Frequency dependency of PBS energy for the overcritical values of a field: 1 — $h = 1$, 2 — $h = 1.001$, 3 — $h = 1.003$. The curve of amplitude is shown for $h = 1.001$.

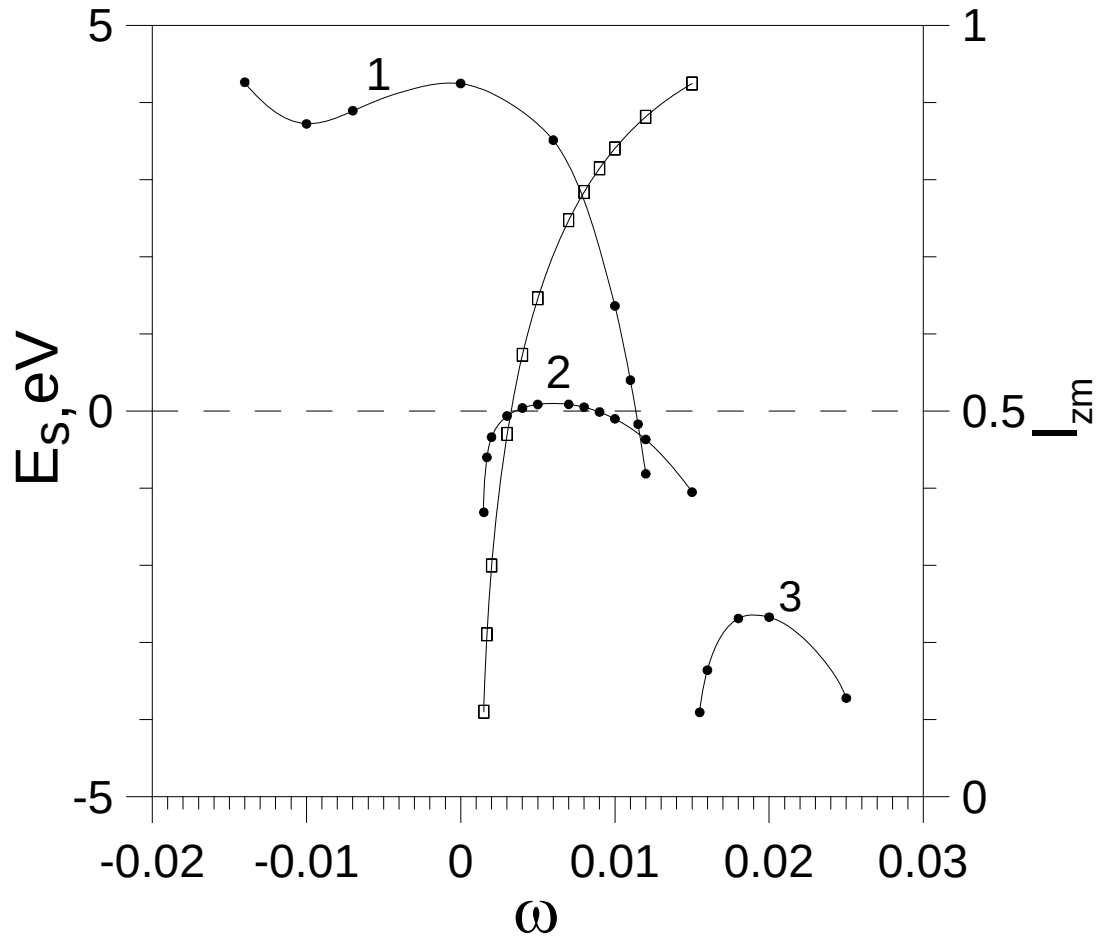

Fig.9 Frequency dependency of PBS energy for reverse spin-flop transition: 1 — $h = 0.91$, 2 — $h = 0.893$, 3 — $h = 0.88$. The curve of amplitude is shown for $h = 0.893$. In this case, the critical value of a field equals $h_{cr} = \sqrt{1 - k_2/k_1} \cong 0.8944$.

In Fig. 10, the frequency dependencies of energy, amplitude, and radius of PBS at $h = 1.001$, $q_m < \sqrt{0.5}$ (it corresponds to curve 2 in Fig. 8) are shown. In this case, corresponding to Eq. (25), the frequency ω decreases in absolute value, and all $q$ values in PBS, for each radius, decrease, i.e. $\omega \to -0.001$ and $q_m \to 0$. However, the energy gets negative and decreases too. Such change of energy can be explained by the fact that simultaneously the radius of PBS increases considerably. The time dependencies of the main parameters of PBS in this case are presented in Fig. 11. Thus, at $h$ value, exceeding the critical value a little, PBS of long duration can originate, which decrease in amplitude and simultaneously increase in radius. Of course, PBS of long duration exist only if they are considered in isolation from other processes. In reality, they will be absorbed by other more quick objects.

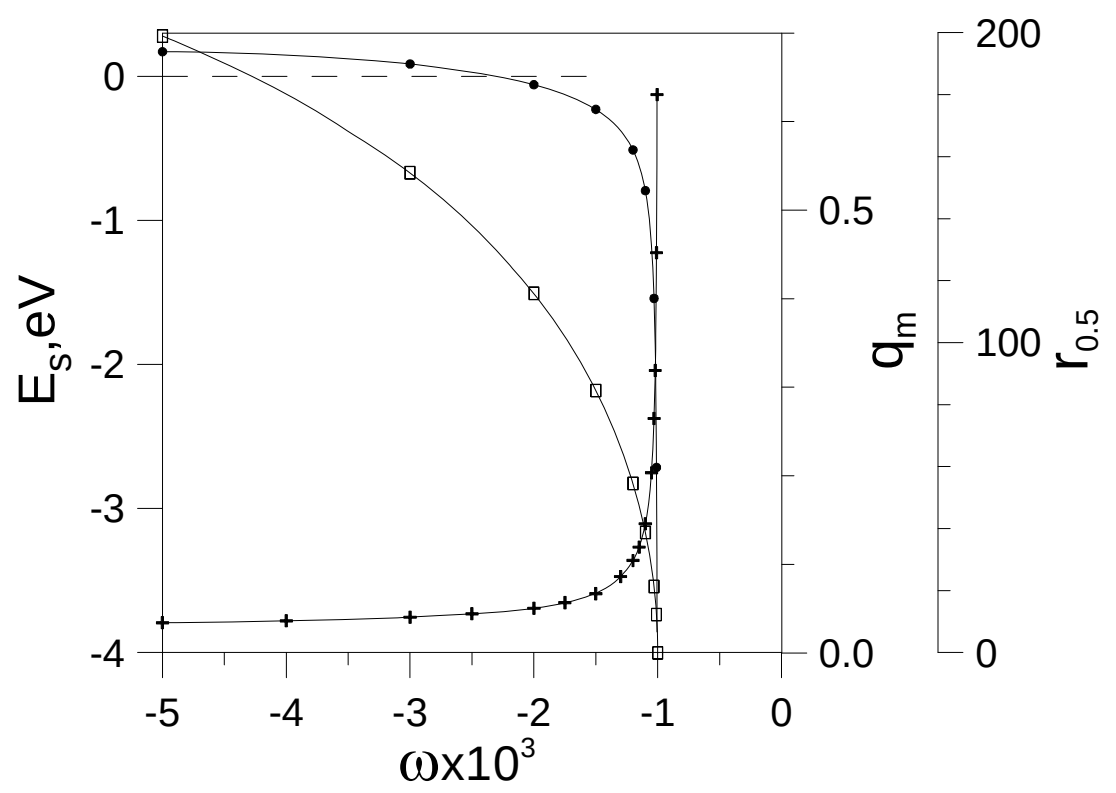

Fig.10 Frequency dependencies of energy, amplitude, and radius of PBS for $h = 1.001$, $q_{m,init} < \sqrt{0.5}$. Here $q_{m,init} = 0.696$.

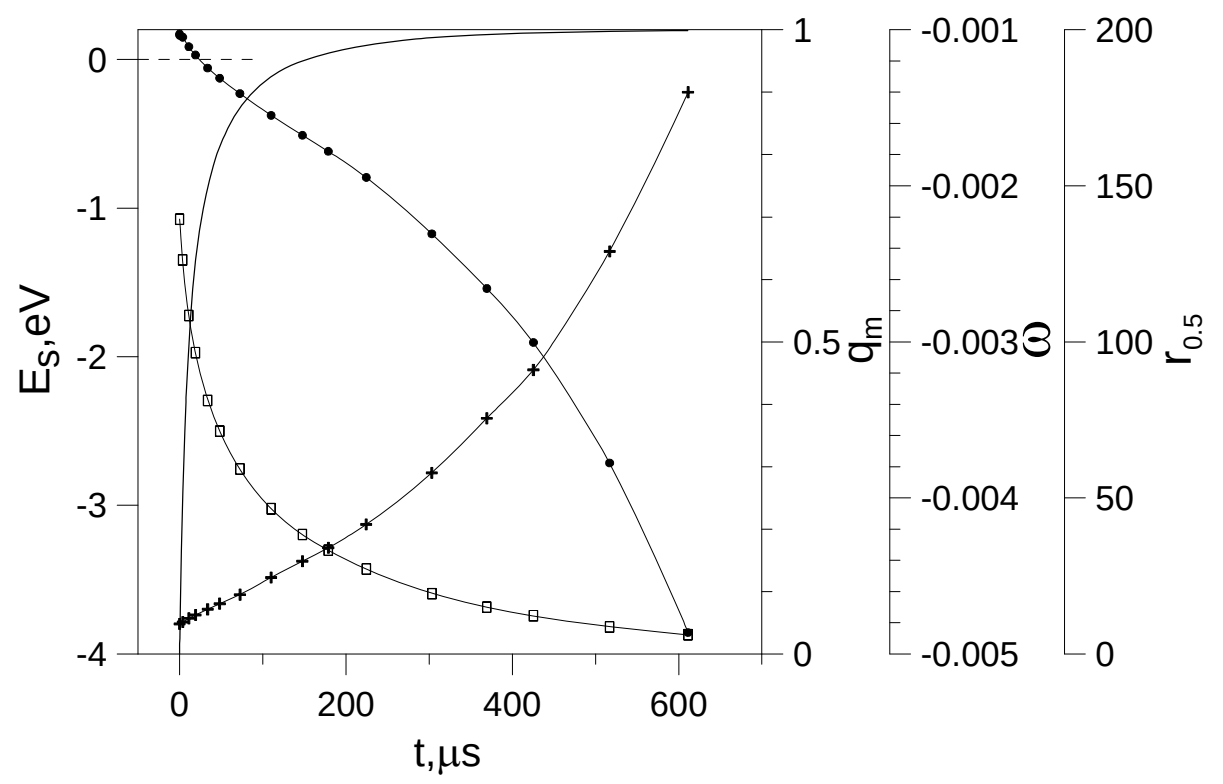

Fig.11 The time dependencies of energy, amplitude, radius, and frequency (continuous line), of PBS if $h = 1.001$ and the initial amplitude $q_{m,init} = 0.696$ (see Fig. 9).

In Fig. 12, the time dependencies of energy, amplitude, and precession frequency at $h = 1.001$, $q_{m,init} > \sqrt{0.5}$ are shown, when the PBS transform into the domains of the high-field phase.

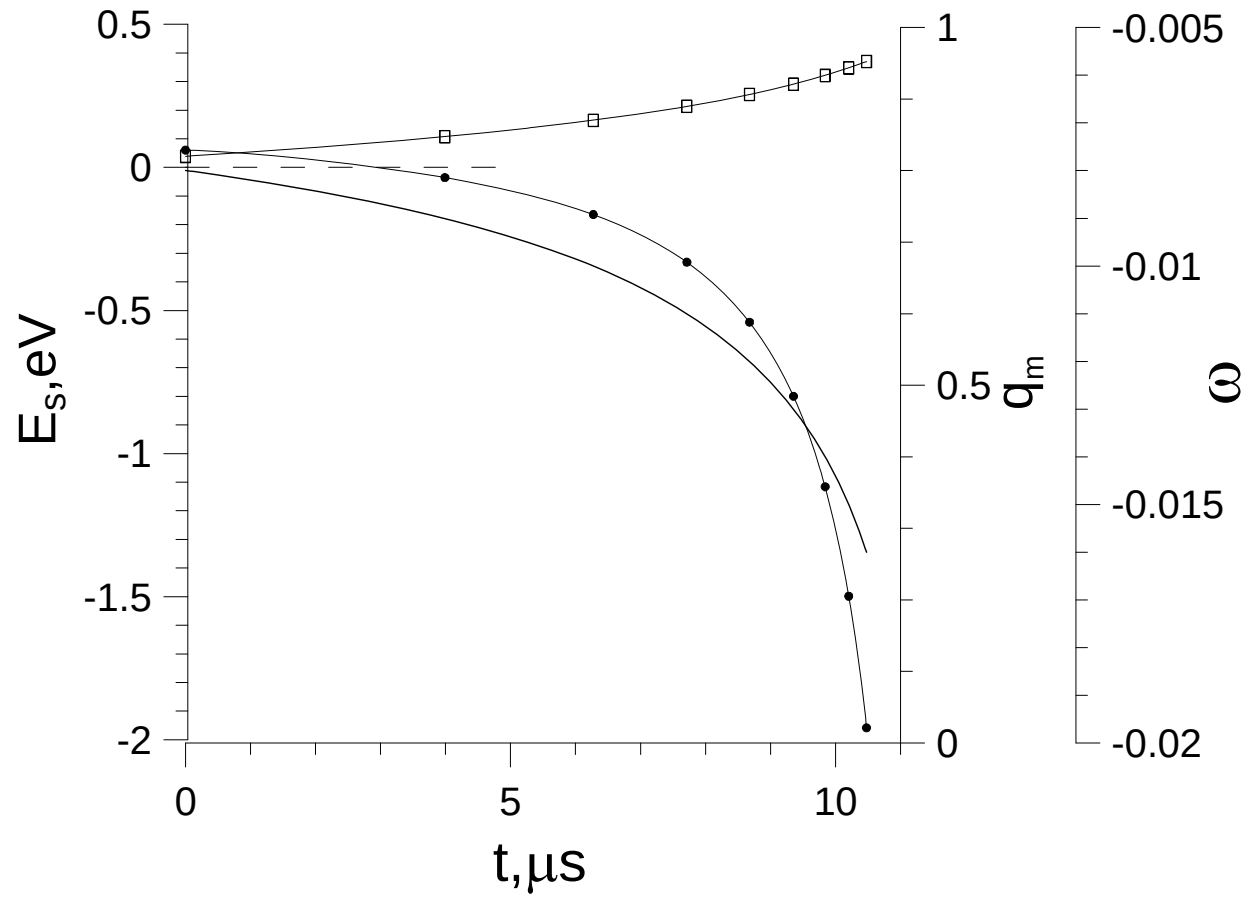

Fig.12 The time dependencies of energy, amplitude, and frequency (continuous line) of PBS if $h = 1.001$ and the initial amplitude $q_{m,init} = 0.82$ (see Fig. 7).

## Conclusions

1. If $h < 1$, the equilibrium PBS of two types are possible: precessing equilibrium solitons for which $\omega > 0$, present the first type; and non-precessing equilibrium solitons for which $\omega = 0$, the second type. In this paper, the process of evolution of PBS into these equilibrium PBS are analyzed. The PBS of the first type are most surprising since in these cases the existence of stable precessing exitations at presence of dissipative member in equation of motion appears to be possible.

**2.** At $h > 1$ (or $h < \sqrt{1 - k_2/k_1}$ for reverse spin-flop transition), the so-called overcritical PBS are possible. It is supposed that such solitons can originate during the disintegration of the initial phase. For such solitons $\omega < 0$ (or $\omega > 0$ for reverse transition). If amplitude of initial PBS $q_{m,init} > \sqrt{0.5}$, such solitons transform into domains of the new phase. The relatively small-amplitude solitons, i.e. when the initial amplitude $q_{m,init} < \sqrt{0.5}$, the PBS decrease slowly to zero in amplitude but increase simultaneously in volume. The time dependencies of energy, amplitude, frequency, and radius of such solitons have been presented. It is possible to interpret such solitons originated during disintegration as "softening" of the initial phase.